\documentclass[
  aps, pra, twocolumn, amsmath, amssymb,
  superscriptaddress, floatfix, nofootinbib
]{revtex4-2}
\usepackage{graphicx}
\usepackage{bm}
\usepackage{xcolor}
\usepackage[colorlinks=true,linkcolor=blue,citecolor=blue,
            urlcolor=blue]{hyperref}
\usepackage{orcidlink}

\newcommand{\ket}[1]{|#1\rangle}
\newcommand{\bra}[1]{\langle#1|}
\newcommand{\Gdec}{\Gamma}
\newcommand{\gdep}{\gamma}
\newcommand{\nhat}{\hat{n}}
\newcommand{\kap}{\kappa}
\newcommand{\PP}{\mathcal{P}}
\newcommand{\Pg}{\mathcal{P}_g}
\newcommand{\Ps}{\mathcal{P}_s}
\newcommand{\ej}{\epsilon_j}
\newcommand{\er}{\epsilon_r}
\newcommand{\ed}{\epsilon_d}
\begin{document}
\title{Decay versus dephasing in Rydberg analog optimization:\\
       the exchange rate depends on the readout}
\author{Seunghyeon Kim\orcidlink{0009-0003-6706-6338}}
\affiliation{North London Collegiate School, Jeju-do 63644, Republic of Korea}
\author{Minjoo Kim\orcidlink{0009-0006-1388-7319}}
\affiliation{Seoul International School, Seongnam-si, Gyeonggi-do 13113, Republic of Korea}
\author{Junwoo Jung\orcidlink{0009-0000-4310-1598}}
\email{gguby@kaist.ac.kr}
\affiliation{Department of Physics, KAIST, Daejeon 34141, Republic of Korea}
\date{\today}
\begin{abstract}
Numerical studies of noisy Rydberg-atom optimization compress decoherence into
a single rate, treating spontaneous decay ($\Gdec$) and dephasing ($\gdep$) as
equivalent. We resolve the two as independent Lindblad axes for quantum
annealing of unit-disk maximum independent set at fixed annealing time and
ask how their relative effect depends on the readout. For the raw
readout the size error collapses onto the effective dephasing rate
$u=\kap\Gdec+\gdep$, and at small noise a unit of decay rate degrades the
result as much as $\kap=7.4$ units of dephasing rate; a collapse fit across
the whole grid returns 8.3. After greedy completion followed by a $(1,2)$-swap local search,
classical post-processing of polynomial cost, the exchange rate falls to
$\kap=1.6$, and at the calibrated operating point of a commercial device
dephasing produces three quarters of the noise-induced error instead of two
fifths. First-order perturbation theory built from noiseless propagation gives
the decay and dephasing susceptibilities of every readout, raw or
post-processed, to a few parts in $10^6$ of the full Lindblad result. The
raw asymmetry is set at the end of the schedule, where a decay event deletes an atom and a dephasing event leaves the readout unchanged. Post-processing refills most
deletions and leaves misplaced atoms, which the two channels produce at
comparable rates. Which channel limits a device therefore depends on the readout as well as on the hardware.
\end{abstract}
\maketitle

\section{Introduction}
\label{sec:intro}
The maximum independent set (MIS) problem asks for the largest set of
vertices of a graph no two of which are adjacent. It is NP-hard even on
unit-disk graphs~\cite{Clark1990}, the class that neutral-atom arrays address
natively. Atoms held in optical tweezers sit at the
vertices~\cite{Barredo2016,Endres2016}, the Rydberg blockade forbids two atoms
within the blockade radius from being excited
together~\cite{Browaeys2020,Bernien2017,Scholl2021}, and the excited atoms at
the end of an anneal therefore form an independent
set~\cite{Pichler2018,Ebadi2022}. Unit-disk MIS has become the standard
benchmark for analog optimization on this
platform~\cite{Ebadi2022,Serret2020,Jeong2025,Wurtz2023,Finzgar2024,Nguyen2023,Andrist2023},
and a body of work asks how hardware noise limits the quality of the returned
sets~\cite{Serret2020,Dalyac2026,Henriet2020,Leclerc2025}.

The Markovian decoherence of the ground--Rydberg qubit has two dominant
channels: decay of the Rydberg state, spontaneous and
blackbody-induced~\cite{Beterov2009}, and dephasing, from scattering off the
intermediate state of the two-photon excitation and from laser phase
noise~\cite{Leseleuc2018,Levine2018}. They are suppressed by different
hardware, a colder environment or a quieter laser, so deciding which to
improve requires their exchange rate $\kap$, the number of units of dephasing
rate that degrade the output as much as one unit of decay rate. Coherence
measurements cannot supply it, because $T_2$ depends on $\Gdec+\gdep$ alone
[Eq.~\eqref{eq:T2}]. Simulations of the platform have not resolved the two
channels independently, so the exchange rate is never measured:
Ref.~\cite{Serret2020} keeps only dephasing, Ref.~\cite{Dalyac2026} varies both
rates by a common factor, which moves a device along a single ray of the
$(\Gdec,\gdep)$ plane, and annealing heuristics inherit the
practice~\cite{Jeong2025,Leclerc2025}. The argument
that dephasing in the energy eigenbasis is benign for adiabatic
computation~\cite{Childs2001,Amin2009,Albash2018RMP} does not settle the
question, because the hardware dephasing operator is diagonal in the
occupation basis, which coincides with the energy eigenbasis only after the
drive is switched off (a related point for constrained annealing is made in
Ref.~\cite{Constrained2024}).

The answer also depends on how the readout is evaluated. Benchmarks report the approximation ratio of the raw readout, while experiments complete each readout into a maximal independent set before evaluating it~\cite{Ebadi2022}, a step that a local search can follow at polynomial cost. A decay event deletes an atom
and leaves a vacancy that such post-processing can refill. A dephasing event
moves no population, and the errors it causes arise through the coherent
dynamics that follow. The two channels can therefore rank differently before
and after post-processing.

Here we resolve decay and dephasing as independent axes at fixed annealing
time and compute their exchange rate for the raw readout and after
post-processing. For the raw readout $\kap\approx7.4$, and decay dominates although the calibrated decay rate of a commercial device is five times
smaller than its dephasing rate. After greedy completion and a $(1,2)$-swap local
search $\kap\approx1.6$, and dephasing produces most of the remaining error.
A first-order theory computed from noiseless propagation reproduces the susceptibilities of every readout. It places the raw asymmetry at the end of the schedule, where decay deletes atoms, and shows that post-processing refills these deletions while leaving misplaced atoms.
Section~\ref{sec:model} defines the model, the errors and the readouts,
Sec.~\ref{sec:map} presents the noise maps, Sec.~\ref{sec:pert} the theory,
Sec.~\ref{sec:robust} its robustness, and Sec.~\ref{sec:using} compares the
two channels at a calibrated operating point.

\section{Model and simulation}
\label{sec:model}

\subsection{Hamiltonian and dissipation}
\label{sec:hamiltonian}
We consider $N$ atoms at positions $\bm{r}_i$, each a two-level system with
ground state $\ket{g}$ and Rydberg state $\ket{r}$, driven by a global Rabi
frequency $\Omega(t)$ and detuning $\delta(t)$~\cite{Browaeys2020,Serret2020},
\begin{equation}
  \hat H(t)=\frac{\Omega(t)}{2}\sum_i\hat\sigma^x_i
  -\delta(t)\sum_i \nhat_i
  +\sum_{i<j}\frac{C_6}{r_{ij}^6}\,\nhat_i\nhat_j ,
  \label{eq:H}
\end{equation}
with $\nhat_i=\ket{r}\bra{r}_i$, $\hbar=1$ and rates in $\mu$s$^{-1}$
(Appendix~\ref{app:ref}). The open dynamics follows the Lindblad
equation~\cite{Lindblad1976,Gorini1976}
\begin{equation}
  \dot\rho=-i[\hat H,\rho]
  +\sum_k \gamma_k\Big(\hat L_k\rho \hat L_k^\dagger
  -\tfrac12\{\hat L_k^\dagger \hat L_k,\rho\}\Big)
  \label{eq:lindblad}
\end{equation}
with two local channels per atom, as in the device models of
Refs.~\cite{Dalyac2026,Serret2020}:
\begin{align}
  \hat L_1^{(i)} &= \ket{g}\bra{r}_i \quad\text{at rate } \Gdec=1/T_1 ,
  \label{eq:L1}\\
  \hat L_2^{(i)} &= \ket{r}\bra{r}_i \quad\text{at rate } \gdep .
  \label{eq:L2}
\end{align}
The single-atom coherence then decays as
\begin{equation}
  1/T_2=(\Gdec+\gdep)/2 ,
  \label{eq:T2}
\end{equation}
so a $T_2$ measurement fixes only the sum. $\Gdec$ collects spontaneous
emission and blackbody depopulation of the Rydberg
level~\cite{Beterov2009,Dalyac2026}; $\gdep$ collects intermediate-state
scattering and laser phase noise~\cite{Leseleuc2018,Levine2018}. A shared
laser dephases the array collectively; Sec.~\ref{sec:collective} treats that
variant.

\subsection{Errors}
\label{sec:errors}
A run returns a bitstring, read as the set $S$ of excited atoms. Let
$\mathcal{M}$ be the set of all $g$ maximum independent sets of the graph and
$|S^*|$ their common size. The approximation ratio and the Hamming distance to
the nearest maximum independent set are
\begin{equation}
  \alpha(S)=\frac{|S|}{|S^*|},\qquad
  d(S)=\min_{S^*\in\mathcal{M}}\big(|S\setminus S^*|+|S^*\setminus S|\big) ,
  \label{eq:alpha}
\end{equation}
where the bracket counts the sites at which the bitstrings of $S$ and $S^*$
differ. For an independent set the distance
splits exactly into two counts,
\begin{equation}
  d(S)=j(S)+2r(S),\quad j=|S^*|-|S|,\quad
  r=\min_{S^*\in\mathcal{M}}|S\setminus S^*| ,
  \label{eq:jr}
\end{equation}
because a nearest solution is reached by removing the $r$ atoms of $S$ it does
not contain and adding $r+j$ missing ones. We call $j$ the \emph{size error}
and $r$ the \emph{placement error}, and normalize both by the MIS size,
$\ej=j/|S^*|=1-\alpha$ and $\er=r/|S^*|$, so that $\ed\equiv d/|S^*|=\ej+2\er$.

The placement error is the part of the Hamming distance that the approximation
ratio does not register. It vanishes on graphs whose maximal independent sets
all have the same size (well-covered graphs~\cite{Plummer1970}); the smallest
graph with placement error is the three-vertex path, whose center atom alone
is maximal but misplaced. Bars denote means over the outcome distribution
$p(S)$ and the graph ensemble.

\subsection{Readouts}
\label{sec:readouts}
Each error is evaluated on three readouts of the same bitstring. The
\emph{raw} readout is $S$ itself. \emph{Greedy completion} $\Pg$ repeatedly
adds the free vertex (no occupied neighbor) with the fewest free neighbors,
ties broken by index, until the set is maximal; this is the post-processing used
in experiments, after blockade violations have been repaired~\cite{Ebadi2022}.
\emph{Swap post-processing} $\Ps$ applies $\Pg$ and then repeats
$(1,2)$-swaps, each removing one atom and adding two mutually non-adjacent
vertices with no other occupied neighbor, in index order until none is
possible, followed by a final completion. The $(1,2)$-swap, also called a
2-improvement, is the standard local-search move for maximum independent
set~\cite{Andrade2012}; a swap step of this kind has been used in place of
greedy post-processing on a Rydberg array~\cite{Jeong2025}. Both run in
polynomial time. We write $\epsilon_x^{\PP}$ for the error of $\PP(S)$ and leave the
readout unlabeled for the raw one. Completion cannot lower the placement error,
since adding atoms never removes a misplaced one; a swap can. An exchange rate
is a ratio of two derivatives of one mean error, so it does not depend on the
normalization by $|S^*|$.

\subsection{Asymmetry of the two channels}
\label{sec:asymmetry}
Both jump operators map independent sets to independent sets. A decay jump
deletes an atom: $j$ rises by exactly one, and $r$ falls by one if the atom
was misplaced and is unchanged otherwise (Appendix~\ref{app:deletion}). A
dephasing jump changes no population. Every error above is diagonal in the
occupation basis, the basis in which $\hat L_2$ is diagonal. Dephasing can therefore increase an error only through the residual dependence of the final error on
individual site occupations, and that dependence vanishes once the drive is
off [Eq.~\eqref{eq:ddepdecomp}]. The eigenbasis argument of
Refs.~\cite{Childs2001,Amin2009} applies only in this regime.

\subsection{Protocol and simulation}
\label{sec:methods}
\emph{Schedule.} We use the three-stage schedule of Ref.~\cite{Serret2020}.
$\Omega$ rises from 0 to $\Omega_0$ during the first $0.25\,t_f$ at fixed
$\delta_0$; $\delta$ is swept linearly from $\delta_0$ to $\delta_{\max}$
during the next $0.44\,t_f$ at fixed $\Omega_0$; $\Omega$ falls back to 0
during the remaining $0.31\,t_f$ at fixed $\delta_{\max}$, with
$\Omega_0/2\pi=1.89$~MHz, $\delta_0/2\pi=-6.00$~MHz,
$\delta_{\max}/2\pi=4.59$~MHz and $C_6/r^6=2\pi\times2.7$~MHz at the unit-disk
distance. The total annealing time is fixed at $t_f=4.56~\mu$s for all noise
strengths. It lies near the optimum at the operating point defined below. A scan at $0.25~\mu$s spacing places that optimum at $3.72~\mu$s, where
$\bar\alpha$ is higher by 0.002 and the first-order exchange rate of the raw
size error differs by 1\%.

\emph{Graphs.} Random unit-disk graphs follow Ref.~\cite{Serret2020}: $N$
vertices uniform in a square of side $\sqrt{N/2}$ with exclusion radius 0.3,
joined when closer than unit distance. The production ensemble has 20
instances at $N=10$ (Appendix~\ref{app:numerics}). Auxiliary ensembles cover
$N=6$--16, 42 site-diluted chain, honeycomb, square, triangular and king's
lattices (the last the geometry of Refs.~\cite{Kim2022,Byun2022,KimData2024})
at $N=12$, and a set of engineered graphs.

\emph{Simulation.} Both jump operators preserve the independent-set subspace,
so the dissipative part of the dynamics restricts to it exactly; the
restriction approximates only the coherent drive, through edges longer than
the distance the final detuning enforces. Against full-space calculations it
raises the raw exchange rate by 11--12\% (Sec.~\ref{sec:subspace},
Appendix~\ref{app:trunc}). The production ensemble is integrated as a density
matrix and larger systems by Monte Carlo wave-function
trajectories~\cite{Dalibard1992,Molmer1993,Plenio1998,Daley2014}.

\emph{Noise grid.} The grid is a $5\times5$ logarithmic lattice extending one
decade on each side of the calibrated operating point of the Pasqal FC1
device~\cite{Dalyac2026}, $\Gdec=0.01~\mu$s$^{-1}$ ($T_1=100~\mu$s) and
$\gdep=0.05~\mu$s$^{-1}$, together with the two single-channel axes: 35
points. The simulations yield the full distribution $p(S)$, from which all means follow exactly, and the three readouts are evaluated on the same distributions.
Uncertainties are bootstrap percentiles (16th and 84th) over graph instances.

\section{Noise maps}
\label{sec:map}
On the dephasing-only axis we reproduce the results of
Ref.~\cite{Serret2020} at the few-percent level (Appendix~\ref{app:ref}).
Without noise the raw approximation ratio is 0.973 at $t_f=4.56~\mu$s, below
one because the anneal is not adiabatic at this duration and because the
ground state of Eq.~\eqref{eq:H} at the end of the sweep is not always a
maximum independent set~\cite{Serret2020}.

\subsection{Raw readout}
\label{sec:collapse}
Plotted against the total rate $\Gdec+\gdep$, the size error does not collapse. At fixed total rate $\bar\alpha$ fans out, ordered by the channel
mixture $\gdep/\Gdec$ [gray points in Fig.~\ref{fig:collapse}(a)]. Allowing one
free parameter, the exchange rate,
\begin{equation}
  u = \kap\,\Gdec + \gdep ,
  \label{eq:u}
\end{equation}
and choosing $\kap$ to minimize the scatter of the 35 points about a common
cubic in $\log u$, fitted to the mean error on a linear scale, gives
\begin{equation}
  \kap_j = 8.33^{+0.52}_{-0.49},
  \label{eq:kappafit}
\end{equation}
with rms residual 0.0033, thirteen times below the value 0.042 for
$\kap=1$ [Fig.~\ref{fig:collapse}(a)]. The residual as a function of the trial
value has a single narrow minimum [Fig.~\ref{fig:collapse}(c)]. The Hamming
distance collapses at $\kap_d=5.15^{+0.59}_{-0.49}$; the placement error
collapses onto $\Gdec+\gdep$ within resolution,
$\kap_r=1.01^{+0.19}_{-0.14}$ (Table~\ref{tab:objectives}).

\subsection{After post-processing}
\label{sec:ppmap}
\begin{figure*}[t]
\centering
\includegraphics[width=\textwidth]{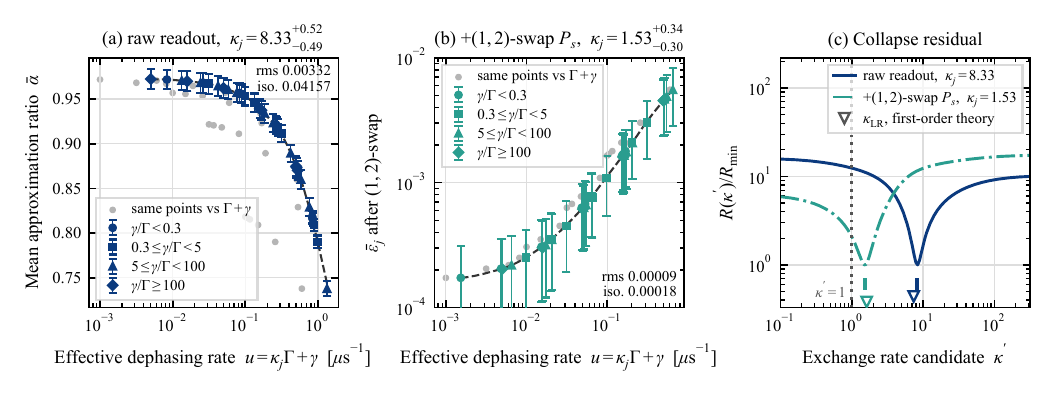}
\caption{\textbf{Noise maps at fixed annealing time} $t_f=4.56~\mu$s. (a) Raw
readout: mean approximation ratio of the 35 grid points against the effective
dephasing rate $u=\kap_j\Gdec+\gdep$ at the best fit $\kap_j=8.33$; marker
shape encodes $\gdep/\Gdec$, and gray points are the same data against
$\Gdec+\gdep$. (b) After greedy completion and $(1,2)$-swap local search: mean
size error on a logarithmic scale, same construction, at the best fit $\kap_j^{\Ps}=1.53$. (c)
Collapse residual against the trial rate $\kap'$, normalized to its minimum;
vertical ticks mark the minima and open triangles the first-order rates
$\kap_{\rm LR}$ of Sec.~\ref{sec:pert}. Dashed curves in (a) and (b), monotone
interpolation through the means in ten bins of $\log u$. Error bars, standard error over 20 graphs.}
\label{fig:collapse}
\end{figure*}

Post-processing removes most of the error. At the FC1 point the mean size
error falls from 0.054 on the raw readout to 0.012 after $\Pg$ and to
0.00076 after $\Ps$, and the probability of returning a maximum independent
set rises from 0.77 to 0.95 and 0.997. Greedy completion leaves the placement
error unchanged (0.0125 to 0.0126); the swap step lowers it eightfold, to
0.0015. The remaining size error still collapses, at
\begin{equation}
  \kap_j^{\Ps}=1.53^{+0.34}_{-0.30},
  \label{eq:kappafitps}
\end{equation}
with rms residual $9\times10^{-5}$ against $1.8\times10^{-4}$ for $\kap=1$
[Fig.~\ref{fig:collapse}(b)]. Post-processing thus lowers the exchange rate by
a factor of five, to a value the greedy completion already reaches on its own
($\kap_j^{\Pg}=1.37$), and leaves an error that the two channels produce at
nearly equal rates.

The first-order theory of Sec.~\ref{sec:pert} gives $\kap_j=7.40$ and
$\kap_j^{\Ps}=1.62$ [triangles in Fig.~\ref{fig:collapse}(c)]. The collapse
values depend on the fitted quantity. The mean error spans three decades after
post-processing, and fitting the same cubic in $\log u$ to its logarithm gives
7.92 and 1.64, with residuals smaller by 26 and 34\% when both fits are measured
on the mean error. Lowering the degree of the polynomial instead raises the
residual by a factor of 3.5 to 10. Across the two fits the
post-processed collapse lies within 6\% of the first-order value and the raw
collapse 7--13\% above it. The
collapse weights the whole grid, which extends a decade beyond the operating
point; the first-order value is the exchange rate at small noise, which the
theory reproduces exactly (Sec.~\ref{sec:lrvalid}).

\section{First-order theory}
\label{sec:pert}

\subsection{Susceptibilities}
\label{sec:damage}
Write the Liouvillian as
$\mathcal{L}=\mathcal{L}_0+\Gdec\mathcal{D}_1+\gdep\mathcal{D}_2$, with
$\mathcal{L}_0$ the coherent part and $\mathcal{D}_k$ the dissipator of
channel $k$ summed over atoms. To first order about the noiseless anneal each
mean error is
\begin{equation}
  \bar\epsilon_x\simeq \epsilon_{x,0} + A_x\,\Gdec + B_x\,\gdep
  =\epsilon_{x,0}+B_x\,u_x ,
  \qquad \kap_x=\frac{A_x}{B_x},
  \label{eq:linexp}
\end{equation}
with $\epsilon_{x,0}$ the noiseless value at the same $t_f$. The decay and
dephasing \emph{susceptibilities} $A_x=\partial\bar\epsilon_x/\partial\Gdec$
and $B_x=\partial\bar\epsilon_x/\partial\gdep$ have units of time. Since the
Hamming distance is linear in the two errors, $A_d=A_j+2A_r$ and
$B_d=B_j+2B_r$. Post-processing is a classical map applied to the bitstring
after measurement. It changes neither the dynamics nor the noise, only the
error assigned to each outcome, $\epsilon(S)\to\epsilon(\PP(S))$, so the same
theory applies to every readout.

Moving the noiseless propagator onto the observable writes each
susceptibility as the time integral of a susceptibility density,
$A_x=\int_0^{t_f}D^x_{\rm dec}\,dt$ and $B_x=\int_0^{t_f}D^x_{\rm dep}\,dt$, with
\begin{align}
  D_{\rm dec}(t) &= -\sum_i\Big[
    \bra{\hat L_1^{(i)}\psi}\hat A(t)\ket{\hat L_1^{(i)}\psi}
    - {\rm Re}\,\bra{\nhat_i\psi}\hat A(t)\ket{\psi}\Big],
  \nonumber\\
  D_{\rm dep}(t) &= -\sum_i\Big[
    \bra{\nhat_i\psi}\hat A(t)\ket{\nhat_i\psi}
    - {\rm Re}\,\bra{\nhat_i\psi}\hat A(t)\ket{\psi}\Big].
  \label{eq:damage}
\end{align}
Here $\ket{\psi(t)}$ is the noiseless state and
$\hat A(t)=\hat U^\dagger(t_f,t)\hat A_f\hat U(t_f,t)$ is the error operator
in the Heisenberg picture, obtained by evolving
$\hat A_f=-\sum_S\epsilon_x(S)\ket{S}\bra{S}$ backward with the noiseless
propagator $\hat U$; its expectation at time $t$ is minus the mean error the
state would reach if the anneal continued without noise. The first term in
each bracket is this expectation after a jump, the second the weight removed
by the non-Hermitian part of the no-jump evolution. Because
$\hat L_1^{(i)\dagger}\hat L_1^{(i)}=\hat L_2^{(i)\dagger}\hat L_2^{(i)}=
\nhat_i$, the second term is the same for both channels, and the anisotropy
lies entirely in what a jump does: decay moves amplitude to the configuration
with atom $i$ removed, dephasing leaves it in place. The split refers to the
standard unraveling with Eqs.~\eqref{eq:L1} and~\eqref{eq:L2} as jump
operators; within a channel only the sum of its two terms is physical. The two
channels stay separately meaningful, since they carry independent rates. This is the adjoint linear
response used in open-system optimal control~\cite{Gautier2025}, here with
every ingredient noiseless.

\subsection{End of the schedule}
\label{sec:deletion}
At $t=t_f$ the error operator is diagonal and commutes with $\nhat_i$, so
$D_{\rm dep}(t_f)=0$ for every error and every readout. The decay
susceptibility density becomes the \emph{terminal deletion response},
\begin{equation}
  D_{\rm dec}(t_f)=\sum_S p_0(S)\sum_{i\in S}
  \big[\epsilon(S\setminus i)-\epsilon(S)\big],
  \label{eq:delresp}
\end{equation}
the mean change of the error when one atom is removed from the returned set,
with $p_0$ the noiseless final distribution. An interval at $\Omega=0$
appended to the schedule adds this amount to $A_x$ per unit time and nothing
to $B_x$. For the raw readout it follows from combinatorics alone
(Appendix~\ref{app:deletion}):
\begin{equation}
  D^{j}_{\rm dec}(t_f)=\alpha_0,\qquad
  D^{r}_{\rm dec}(t_f)=-\,\mathbb{E}_{p_0}\!\left[\frac{w(S)}{|S^*|}\right]
  \le 0 ,
  \label{eq:termjr}
\end{equation}
where $w(S)\ge r(S)$ counts the atoms of $S$ absent from at least one nearest
maximum set. Removing an atom at the end always raises $j$ by one and never
raises $r$.

After post-processing the size error responds differently. Consider an output
that post-processing leaves unchanged. A deletion opens a vacancy, which
completion refills with the removed atom or with one or more of its neighbors,
and the swap step never shrinks a set. The deletion response of the
post-processed size error is therefore zero or negative on such outputs
(Appendix~\ref{app:deletion}). Only outputs that post-processing changes can
contribute positive terms. The measured responses are $\alpha_0=0.973$ for the
raw size error, $-0.0066$ after $\Pg$ and $+0.0017$ after $\Ps$
[Fig.~\ref{fig:damage}(a), Table~\ref{tab:objectives}].

\subsection{Validation and the single atom}
\label{sec:lrvalid}
Equation~\eqref{eq:damage} requires one forward and one backward noiseless
propagation per instance. Against finite differences of the full Lindblad
solver at the same $t_f$, the eighteen susceptibilities of
Table~\ref{tab:objectives} (decay and dephasing, three errors, three
readouts) agree to within $2.2\times10^{-6}$ in relative terms
[Fig.~\ref{fig:validation}(a)]. A single atom driven by the same schedule
already has $\kap_j=8.53$ on the raw readout, and a blockaded pair 11.97, so
the raw asymmetry is single-atom physics that interactions modulate by tens of
percent. The placement error and the post-processed errors vanish for one or
two atoms and are many-body quantities by construction.

\subsection{How post-processing changes the exchange rate}
\label{sec:completion}
\begin{figure*}[t]
\centering
\includegraphics[width=\textwidth]{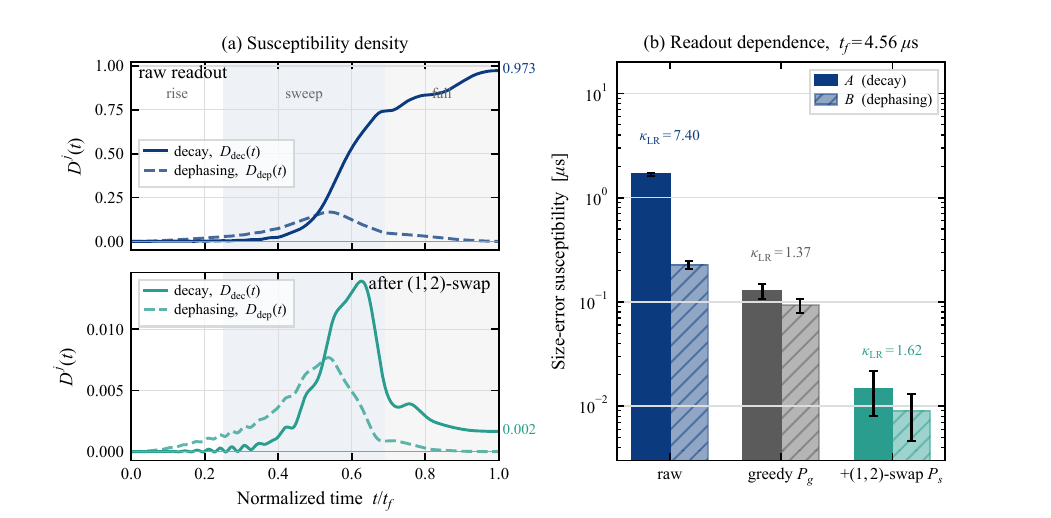}
\caption{\textbf{Mechanism.} (a) Susceptibility densities of
Eq.~\eqref{eq:damage} for the size error of the raw readout (top) and after
swap post-processing (bottom), averaged over the 20 production instances
at $t_f=4.56~\mu$s; the areas are the susceptibilities. The dephasing density
vanishes once the drive is off; the end values are the terminal deletion
responses of Eq.~\eqref{eq:delresp}. (b) Decay ($A$) and dephasing ($B$)
susceptibilities of the size error for the three readouts, with their ratio
$\kap_{\rm LR}$. Error bars, bootstrap 68\% over instances.}
\label{fig:damage}
\end{figure*}

Post-processing lowers both susceptibilities of the size error, by different
factors [Fig.~\ref{fig:damage}(b)]. From the raw readout to $\Ps$, $A_j$ falls
from 1.68 to $0.0147~\mu$s, a factor of 114, and $B_j$ from 0.227 to
$0.0090~\mu$s, a factor of 25. The susceptibility densities show where the
difference comes from [Fig.~\ref{fig:damage}(a)]. On the raw readout the decay
density keeps rising through the ramp-down and reaches $\alpha_0$ at the end,
where a decay event simply deletes an atom. After $\Ps$ it peaks near the end of
the sweep and falls to 0.002 at the end, because post-processing refills the late deletions and leaves misplaced atoms. After $\Ps$ the placement error exceeds the size error ($\bar\epsilon_r^{\Ps}=0.0015$ against
$\bar\epsilon_j^{\Ps}=0.0008$ at FC1), and the raw placement error already has
an exchange rate near one, since per unit rate the two channels misplace atoms almost equally ($A_r=0.135$, $B_r=0.099~\mu$s, $\kap_r=1.37$). The
post-processed errors have rates of the same size, 1.37 after $\Pg$ and 1.62
after $\Ps$. A bootstrap over instances puts the difference at
$+0.25^{+0.39}_{-0.35}$, so the two post-processing rules cannot be told apart
by their exchange rate. What the swap step changes is the size of the error,
not its anisotropy: at the FC1 point it lowers $\bar\ej$ by a factor of 16 and
$1-p_1$ by the same factor. The collapse of $\kap_j$ is therefore a property of
refilling deletions rather than of one algorithm.

\begin{table*}[t]
\caption{Susceptibilities and exchange rates at $t_f=4.56~\mu$s on the 20
production instances. $D_{\rm dec}(t_f)$, terminal deletion response; $A$, $B$
in $\mu$s; $\kap_{\rm LR}=A/B$; $\kap_{\rm FD}$, finite differences of the full
Lindblad solver at the same $t_f$; $\kap_{\rm coll}$, collapse fit over the
noise grid at the same $t_f$; last two columns, mean error without noise and
at the FC1 point.}
\label{tab:objectives}
\begin{ruledtabular}
\begin{tabular}{llcccccccc}
readout & error & $D_{\rm dec}(t_f)$ & $A$ & $B$ & $\kap_{\rm LR}$ & $\kap_{\rm FD}$ & $\kap_{\rm coll}$ & noiseless & FC1 \\
\hline
raw & size $\ej$ & $+0.973$ & 1.681 & 0.227 & 7.40 & 7.40 & $8.33^{+0.52}_{-0.49}$ & 0.0266 & 0.0542 \\
 & placement $\er$ & $-0.0069$ & 0.135 & 0.099 & 1.37 & 1.37 & $1.01^{+0.19}_{-0.14}$ & 0.0065 & 0.0125 \\
 & Hamming $\ed$ & $+0.960$ & 1.951 & 0.425 & 4.60 & 4.60 & $5.15^{+0.59}_{-0.49}$ & 0.0396 & 0.0792 \\
\hline
greedy $\Pg$ & size & $-0.0066$ & 0.127 & 0.093 & 1.37 & 1.37 & $1.05^{+0.19}_{-0.14}$ & 0.0064 & 0.0120 \\
\hline
$+(1,2)$-swap $\Ps$ & size & $+0.0017$ & 0.0147 & 0.0090 & 1.62 & 1.62 & $1.53^{+0.34}_{-0.30}$ & 0.00016 & 0.00076 \\
 & placement & $+0.0034$ & 0.0300 & 0.0183 & 1.64 & 1.64 & $1.54^{+0.33}_{-0.30}$ & 0.00032 & 0.00155 \\
 & Hamming & $+0.0084$ & 0.0747 & 0.0456 & 1.64 & 1.64 & $1.54^{+0.34}_{-0.30}$ & 0.00079 & 0.00386 \\
\end{tabular}
\end{ruledtabular}
\end{table*}

\subsection{Origin of the dephasing susceptibility}
\label{sec:Bvar}
Across 122 instances (random ensembles at $N=6$--16 and the site-diluted
lattices), the raw decay susceptibility is nearly a constant of the protocol,
$A_j=1.79\pm0.16~\mu$s, while $B_j$ varies severalfold, so that the
per-instance $\kap_j$ reaches 16. $B_j$ does not follow the coherence of the
noiseless trajectory. The integrated single-site variance
$\int\sum_i\langle\nhat_i\rangle(1-\langle\nhat_i\rangle)dt$, which sets the
purity lost to local dephasing, is uncorrelated with it ($r=-0.15$). The MIS
degeneracy $g$ and the variance of $\alpha$ over the final distribution are
correlated with it (Appendix~\ref{app:Bvar}). The reason follows from Eq.~\eqref{eq:damage}. The expectation of the error operator is constant in
time, $\bra{\psi}\hat A(t)\ket{\psi}=-\epsilon_{x,0}$, and subtracting it,
$\hat A=-\epsilon_{x,0}+\delta\hat A$, leaves
\begin{equation}
  D_{\rm dep}(t)=\sum_i\Big[{\rm Re}\,\langle\nhat_i\,\delta\hat A\rangle
  -\langle\nhat_i\,\delta\hat A\,\nhat_i\rangle\Big].
  \label{eq:ddepdecomp}
\end{equation}
The dephasing contribution is set by how strongly the final error still
depends on individual site occupations. It can be small for a strongly
delocalized state whose branches have nearly the same error, as on a graph
with many maximum independent sets. Since no simple quantity predicts $B_j$,
the exchange rate has to be computed for each problem class, which requires
only noiseless propagation.

\subsection{Dependence on the schedule}
\label{sec:fall}
Since dephasing loses its effect as the drive is switched off while a decay
event at the end still raises the raw size error, the exchange rate depends on
the schedule. The rates quoted here refer to the schedule of
Ref.~\cite{Serret2020}, which spends 31\% of the anneal ramping $\Omega$ down,
and change with the schedule.

\section{Robustness}
\label{sec:robust}

\subsection{System size}
\label{sec:size}
On the raw readout the first-order rate is 7.93, 7.74, 7.16, 6.81, 5.96 and
7.20 at $N=6$, 8, 10, 12, 14 and 16 [Fig.~\ref{fig:geometry}(a)], close to the
single-atom value. These are the auxiliary size-scan ensembles of
Sec.~\ref{sec:methods}; the production ensemble, a different draw of 20 graphs
at $N=10$, gives 7.40 (Table~\ref{tab:objectives}), and the spread between the
two is the instance-to-instance scatter at fixed $N$. After $\Ps$ it is 3.78 and 3.13 at $N=6$ and 8 and between
1.1 and 1.8 for $N=10$--16. These values rest on fewer graphs than the raw
ones. On the smallest graphs post-processing returns a maximum independent set
in nearly every run, and only 2 of the 20 graphs at $N=6$ and 3 of the 16 at
$N=8$ have a nonzero post-processed error; the others have $A=B=0$ and drop out
of the ratio. At $N=16$ one of the eight graphs has negative post-processed
susceptibilities, larger in magnitude than those of the other seven combined,
and the value quoted leaves it out. The noiseless post-processed size error is
$10^{-6}$ at $N=6$ and rises with $N$, to $4\times10^{-5}$,
$2.0\times10^{-4}$ and $3.2\times10^{-4}$ at $N=8$, 10 and 12.

\subsection{Degeneracy and connectivity}
\label{sec:geometry}
The 42 site-diluted lattices at $N=12$ span realized mean degrees
$\langle k\rangle=1.0$--4.8 and MIS degeneracies $g=1$--24. The raw rate rises
with $g$ [Fig.~\ref{fig:geometry}(b)]. Against the degree alone it falls,
$d\kap_j/d\langle k\rangle=-0.42\pm0.29$ (Fig.~\ref{fig:degree}), because degree
and $\log g$ are anticorrelated in this set ($r=-0.21$); the joint fit of
Eq.~\eqref{eq:condfit} gives $a=0.19\pm0.02$ and $b=-0.006\pm0.013$: at fixed
degeneracy the degree has no resolved effect, bounded by $|b|<0.02$ at one
standard error. Engineered graphs extend the range
(Table~\ref{tab:specials}). Even cycles ($g=2$) give 7.5--7.9 and odd cycles
($g=n$) 14.3--15.9; complete graphs give the largest rates, 17.4 for $K_4$ to
26.1 for $K_8$; and removing a single edge from $K_5$, which makes the maximum
set unique, lowers the rate from 19.7 to 4.4 at nearly unchanged
connectivity. The post-processed rates spread between about 0 and 6 without a
systematic dependence on $g$.

\subsection{Collective dephasing}
\label{sec:collective}
\begin{figure*}[t]
\centering
\includegraphics[width=\textwidth]{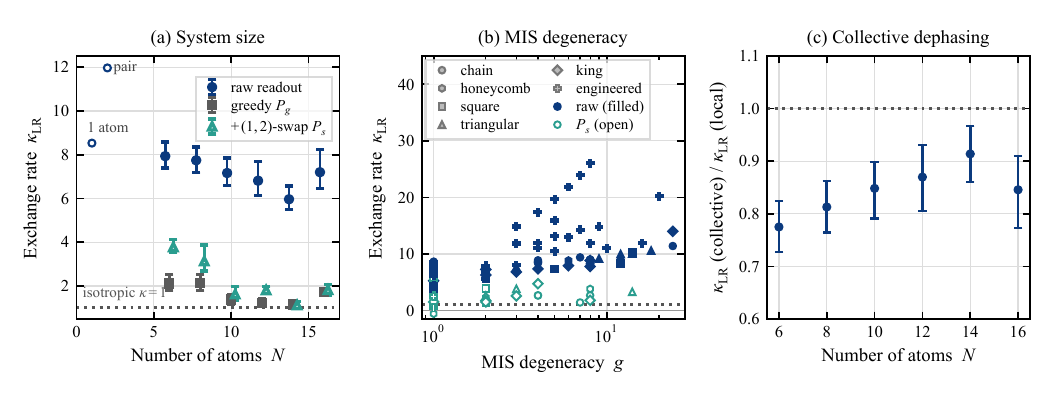}
\caption{\textbf{Robustness.} First-order exchange rates of the size error at
$t_f=4.56~\mu$s. (a) System size, for the raw readout (blue), after greedy
completion $\Pg$ (gray) and after a further $(1,2)$-swap local search $\Ps$
(teal); $N=1$ and 2 are a single atom and a blockaded pair, raw readout only.
The greedy series is computed from the same noiseless propagations as the
others. Post-processed points at $N=6$ and 8 rest on 2 and 3 graphs, and at
$N=16$ one graph is left out (Sec.~\ref{sec:size}). (b) MIS degeneracy, raw readout and $\Ps$, for 42 site-diluted lattices at $N=12$
(marker shape by family) and the engineered graphs of
Table~\ref{tab:specials} (crosses); filled, raw; open, post-processed. (c) Ratio of
the raw-readout rate under collective dephasing to that under local dephasing,
with rates matched to the same $T_2$. Error bars, bootstrap 68\% over
instances, paired in (c).}
\label{fig:geometry}
\end{figure*}

Atoms driven by one laser share its phase noise, for which the appropriate jump operator is the collective $\hat L=\sum_i\nhat_i$. With its rate matched to
the same single-atom $T_2$, it raises $B_j$ from 0.227 to $0.274~\mu$s and
lowers the raw rate from 7.40 to 6.12 on the production ensemble; the cross
terms $i\neq j$ of the collective dissipator couple the error to pairs of
sites, which is the direction Eq.~\eqref{eq:ddepdecomp} predicts. Across
$N=6$--16 collective dephasing lowers the raw rate by a factor between 0.78
and 0.91, with the 68\% interval below one at every size
[Fig.~\ref{fig:geometry}(c)]. For the post-processed readout the same ratio
ranges from 0.6 to 1.3 without a consistent sign, which the few graphs with
nonzero post-processed error (Sec.~\ref{sec:size}) do not resolve. The local model is therefore the optimistic one
for the raw readout.

\subsection{Subspace restriction}
\label{sec:subspace}
Full-space calculations on the same graphs give a raw rate of 7.05 against
7.93 in the subspace at $N=6$ and 6.79 against 7.74 at $N=8$, lower by 11 and
12\% (Appendix~\ref{app:trunc}). For the post-processed size error the
full-space rates are 3.48 and 3.30 against 3.78 and 3.13, a difference within
the uncertainties.

Three determinations of $\kap_j$ are collected in Fig.~\ref{fig:validation}(b).
The collapse fit of Sec.~\ref{sec:collapse} weights the whole noise grid. Axis
matching uses only the two single-channel axes: it interpolates each axis onto
the other to find the pair of rates that produce the same mean error, and takes
the geometric mean of the ratios obtained in the two directions. First-order
theory gives $A/B$ at vanishing noise. The three give 8.33, 7.58 and 7.40 for
the raw readout and 1.53, 1.63 and 1.62 after $\Ps$.
We quote the first-order values, $\kap_j=7.40$ for the raw
readout and $\kap_j^{\Ps}=1.62$ after swap post-processing, with the
statistical uncertainties of Fig.~\ref{fig:validation}(b) and a systematic
shift of $-12\%$ for the raw value from the subspace restriction.

\section{Relative contribution of the two channels}
\label{sec:using}
At the FC1 point the effective dephasing rate of the raw size error is
$u_j=\kap_j\Gdec+\gdep=0.124~\mu$s$^{-1}$. Decay accounts for 60\% of the
Markovian contribution, although its rate is five times smaller than the
dephasing rate (Table~\ref{tab:share}). After swap post-processing
$u_j^{\Ps}=0.066~\mu$s$^{-1}$ and the decay share of that contribution falls to
24\%. These are shares of the term that the noise adds, $A\Gdec+B\gdep$, not of
the total error: the noiseless anneal already contributes $\epsilon_{j,0}=0.027$
of the raw error 0.054 at this point, and 0.00016 of the 0.00076 left after
$\Ps$.

The two channels are therefore worth different amounts of engineering effort,
and the ranking reverses with the readout. On the raw readout, halving the decay
rate lowers $u_j$ by $0.037~\mu$s$^{-1}$ and halving the dephasing rate by
$0.025~\mu$s$^{-1}$. After post-processing the same two changes lower
$u_j^{\Ps}$ by 0.008 and $0.025~\mu$s$^{-1}$: halving the dephasing rate has
become three times as effective as halving the decay rate.

The share also depends on how much of the measured decay rate deletes atoms.
The measured $1/T_1$ includes blackbody transfer to neighboring Rydberg levels,
which our $\hat L_1$ does not describe; only the fraction $b$ of decays that
returns the atom to $\ket{g}$ acts as a deletion. Rb $60S_{1/2}$ at 300~K has a computed lifetime that
reproduces the calibrated $T_1=100~\mu$s, and for it ARC~\cite{Sibalic2017}
gives $b=0.44$, counting the zero-temperature radiative decays that cascade to
the ground state within the schedule. The value is insensitive to the level
cutoff of the sum: $b=0.434$--$0.435$ for cutoffs between $n=75$ and $n=85$. Counting only these decays, the decay share is 39\% on the raw
readout and 12\% after post-processing (Table~\ref{tab:share}). The last column of Table~\ref{tab:share} is therefore conditional on one decay
path. The transferred atoms form a third error channel that leaves the
computational subspace, carries its own effect on the readout, and is not
modeled here.

\begin{table}[t]
\caption{Decay share of the Markovian contribution at the FC1 point
($\Gdec=0.01$, $\gdep=0.05~\mu$s$^{-1}$), from the first-order rates of
Table~\ref{tab:objectives}; last column, counting only the fraction $b=0.44$
of decays that return the atom to $\ket{g}$.}
\label{tab:share}
\begin{ruledtabular}
\begin{tabular}{lcccc}
readout, error & $\kap_{\rm LR}$ & $u$ ($\mu$s$^{-1}$) & decay share & with $b$ \\
\hline
raw, size & 7.40 & 0.124 & 60\% & 39\% \\
raw, Hamming & 4.60 & 0.096 & 48\% & 29\% \\
greedy $\Pg$, size & 1.37 & 0.064 & 22\% & 11\% \\
swap $\Ps$, size & 1.62 & 0.066 & 24\% & 12\% \\
\end{tabular}
\end{ruledtabular}
\end{table}

\emph{Scope.} Scaling both rates by a common factor~\cite{Dalyac2026} moves a
device along a collapse curve and remains valid for a given readout.
Comparing devices by their total rate is not valid, since devices with equal
$\Gdec+\gdep$ return different approximation ratios [gray points in
Fig.~\ref{fig:collapse}(a)]. Any combinatorial objective read from a
bitstring is diagonal in the basis of the dephasing operator, whether it is
reached variationally~\cite{Farhi2014} or by annealing. Dephasing therefore has
no effect once the drive is off. Whether decay still has an
effect then depends on the sign of the deletion response,
Eq.~\eqref{eq:delresp}, which is positive for the raw size error and close to
zero after post-processing. A large exchange rate thus requires an error that increases when a single atom is deleted, and a noiseless
calculation decides which case applies. Of the classical noise sources, quasistatic
detuning can be placed on the same axis; adiabaticity suppresses it by three
orders of magnitude relative to the equivalent Lindblad rate
(Appendix~\ref{app:classical}).

\section{Conclusion}
\label{sec:discussion}
In Rydberg analog optimization the exchange rate between decay and dephasing
depends on the readout. On the raw readout a unit of decay rate degrades the result as much as 7.4
units of dephasing rate, and at the calibrated operating point of a commercial
device it produces 60\% of the noise-induced error. After greedy completion and
a $(1,2)$-swap local search the exchange rate is 1.6, and dephasing produces
three quarters of what the noise still costs. A first-order theory built from noiseless propagation reproduces the
susceptibilities of every readout to a few parts in $10^6$. It locates the raw
asymmetry at the end of the schedule, where a dephasing event leaves the
readout unchanged and a decay event deletes one atom, and it shows that
post-processing refills these deletions while leaving misplaced atoms, which
both channels produce at comparable rates. A single effective rate $u$
therefore describes a device only for a given readout. Because the
susceptibilities are noiseless quantities, they could also serve as objectives
for schedule design; we do not pursue that here.

The local dephasing model is the optimistic one for the raw readout;
collective dephasing lowers the raw rate by 17\% at $N=10$. The
independent-set subspace raises the raw rate by 11--12\% against full-space
calculations at $N=6$ and 8. Blackbody transfer to neighboring Rydberg levels
is a third error channel that leaves the computational subspace and is not
modeled. Among classical noise sources only quasistatic detuning has been tested,
and post-processing is represented by one scheme. The exchange rates are
computed for $N\le16$ and the post-processed rate depends on $N$ below
$N=10$.

\begin{acknowledgments}
This work received no external funding.
\end{acknowledgments}

\section*{Data availability}
The code, random seeds and outcome distributions $p(S)$ from which every
quantity in this work can be recomputed are openly available at
Zenodo~\cite{zenodo}.

\appendix

\section{Numerical methods}
\label{app:numerics}
\emph{Density-matrix solver.} The production grid integrates
Eq.~\eqref{eq:lindblad} on the independent-set subspace. Its stiffness,
diagonal Liouvillian entries of order $\delta_{\max}N\sim250~\mu$s$^{-1}$
against a Rabi scale of $12~\mu$s$^{-1}$, is removed by an integrating factor:
the part diagonal in the matrix-element basis (detuning, interaction and all
damping) is integrated exactly, and only the Rabi commutator and the decay
gain term are advanced by fourth-order Runge--Kutta with $dt=0.005~\mu$s.
Halving or quartering $dt$ changes the post-processed error at FC1 by less
than $10^{-9}$, and no negative probabilities occur.

\emph{Trajectories.} Because $\hat L_1^{(i)\dagger}\hat L_1^{(i)}=
\hat L_2^{(i)\dagger}\hat L_2^{(i)}=\nhat_i$, the effective Hamiltonian
$\hat H-\tfrac{i}{2}(\Gdec+\gdep)\sum_i\nhat_i$ of the Monte Carlo
wave-function method is diagonal in its non-Hermitian part and fits the same
integrating factor. A decay jump moves each configuration containing $i$ to
the one with $i$ removed; a dephasing jump projects onto configurations
containing $i$.

\emph{First-order quantities.} The susceptibilities are evaluated from
Eq.~\eqref{eq:damage} by propagating the noiseless state forward and the error
operator backward on the subspace. In the full space the interaction energy of
blockade-violating configurations, of order $C_6/0.3^6\approx2\times10^4~\mu$s$^{-1}$,
makes this route unstable at the step size above, and the full-space rates of
Sec.~\ref{sec:subspace} are finite differences of the full Lindblad solver.

\emph{Tests.} The implementation reproduces the single-atom laws
$\rho_{rr}=e^{-\Gdec t}$ and $|\rho_{gr}|=\tfrac12e^{-(\Gdec+\gdep)t/2}$ to
$10^{-9}$ and preserves trace and Hermiticity.

\emph{Instances.} The production graphs are generated with seeds 0--19. Their
subspace dimensions are 152, 86, 56, 87, 66, 95, 84, 87, 55, 69, 72, 94, 72,
120, 34, 46, 93, 92, 106, 77, their MIS sizes 6, 4, 4, 4, 4, 5, 4, 4, 4, 3, 4,
5, 4, 5, 3, 4, 4, 5, 4, 4, and their MIS degeneracies 2, 8, 3, 12, 4, 2, 10,
12, 1, 28, 5, 2, 5, 4, 5, 1, 12, 2, 19, 8. Seeds 9 and 18 are well-covered,
so their placement error vanishes identically.

\begin{figure*}[t]
\centering
\includegraphics[width=\textwidth]{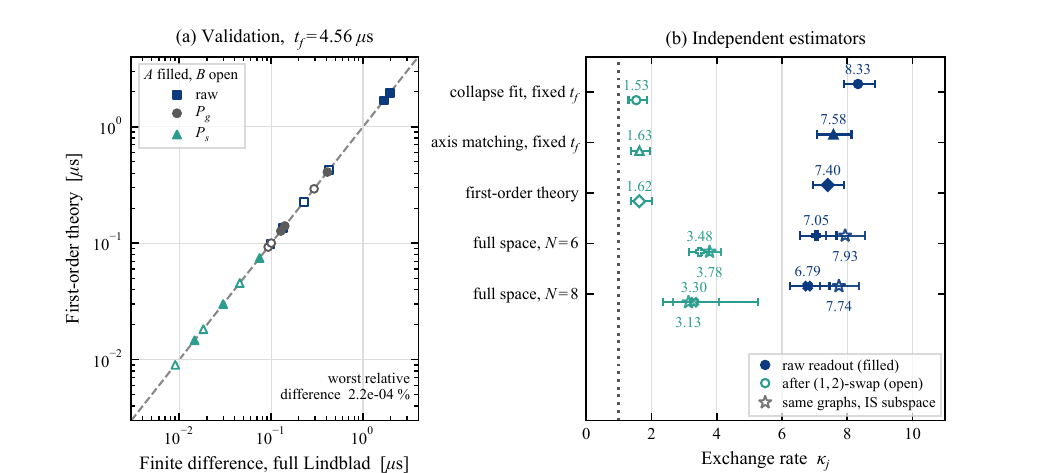}
\caption{\textbf{Validation.} (a) First-order susceptibilities
against finite differences of the full Lindblad solver at $t_f=4.56~\mu$s for
the size, placement and Hamming errors of the three readouts; filled, decay
($A$); open, dephasing ($B$). (b) Independent determinations of the exchange
rate of the size error for the raw readout (filled) and after swap
post-processing (open): collapse fit and axis matching at fixed $t_f$ and
first-order theory on the production ensemble, and full-space finite
differences at $N=6$ and 8, with the subspace values for the same graphs
(stars). Error bars, bootstrap 68\% over instances.}
\label{fig:validation}
\end{figure*}

\section{Subspace restriction}
\label{app:trunc}
Write $V(r)=C_6/r^6$ for the interaction of Eq.~\eqref{eq:H}. An edge is enforced at
the end of the sweep only if its interaction exceeds $\delta_{\max}$, i.e.\ for distances
below $(V(1)/\delta_{\max})^{1/6}=0.915$.
About 15\% of edges in the random ensemble are longer, and in some instances
the ground state of the final Hamiltonian is not an independent
set~\cite{Serret2020,Pichler2018b}. The full-space calculation keeps all
$2^N$ configurations and repairs each blockade-violating bitstring to its
largest independent subset before evaluation, as experimental
post-processing does~\cite{Ebadi2022}.

On the same graphs and at the same $t_f$ the raw rate is $7.05$ [6.52, 7.69] in
the full space against $7.93$ [7.38, 8.58] in the subspace at $N=6$ (20 graphs).
At $N=8$ (16 graphs) it is $6.79$ [6.27, 7.45] against $7.74$ [7.18, 8.37].
After swap post-processing the full-space rates are 3.48 [3.15, 3.77] and
3.30 [2.35, 5.25] against 3.78 [3.53, 4.12] and 3.13 [2.64, 4.05]. The noiseless
final state leaks out of the subspace with median weight $8.5\times10^{-5}$ at
$N=6$ and $5\times10^{-3}$ at $N=8$; on 2 graphs at $N=6$ and 1 at $N=8$ it
leaks by more than one half.

\section{Units and reference results}
\label{app:ref}
Both reference studies quote Lindblad rates in angular-frequency units:
Ref.~\cite{Serret2020} states that $\gdep=3.0$ corresponds to
$\gdep/2\pi=0.48$~MHz, and the decay rate $0.01~\mu$s$^{-1}$ of
Ref.~\cite{Dalyac2026} matches their computed $n=60$ lifetime of
$100.6~\mu$s~\cite{Beterov2009}. Choosing the annealing time on a
nine-point logarithmic grid, as Ref.~\cite{Serret2020} does, we obtain on the
dephasing-only axis $\bar{\alpha}=0.767\pm0.014$ for $\gdep=3.0~\mu$s$^{-1}$
at $t_f\approx1.3~\mu$s, against about 0.78 at 1.3--1.5~$\mu$s in
Ref.~\cite{Serret2020}, and 0.916 for $\gdep=0.3~\mu$s$^{-1}$ at
$t_f\approx3.2~\mu$s, against about 0.93 at $3~\mu$s. Such a grid locates the
optimum only coarsely. At FC1 it places the optimum at $4.56~\mu$s, while a
scan at $0.25~\mu$s spacing places it at $3.72~\mu$s, with $\bar\alpha=0.946$
and 0.948 at the two times.

\section{Dephasing susceptibility and connectivity}
\label{app:Bvar}
\begin{figure*}[t]
\centering
\includegraphics[width=0.9\textwidth]{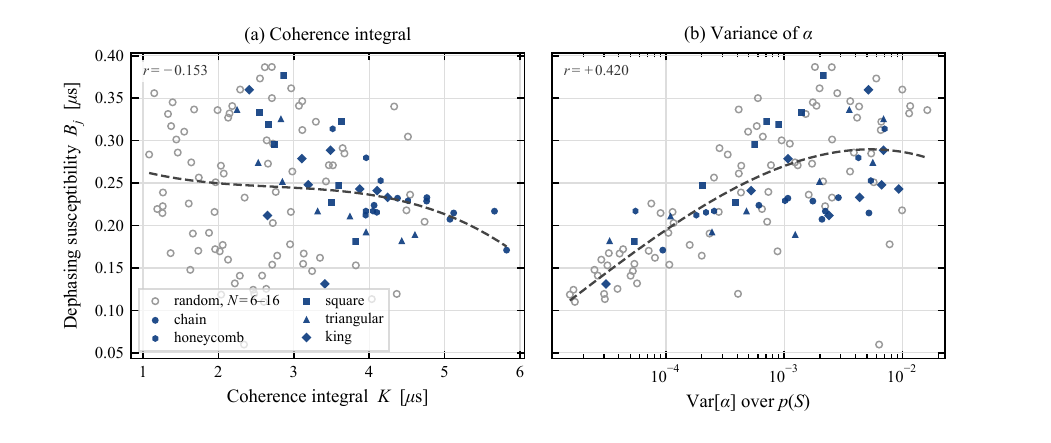}
\caption{\textbf{Origin of the dephasing susceptibility.} Raw $B_j$ for 122
instances (random ensembles at $N=6$--16, open gray; site-diluted lattices,
filled) against (a) the integrated single-site variance $K$ of the noiseless
trajectory and (b) the variance of $\alpha$ over the final distribution; $r$,
correlation coefficient; dashed, best cubic in the logarithm of the
abscissa.}
\label{fig:mechanism}
\end{figure*}

\emph{Correlates of $B_j$.} For local dephasing of a pure state the purity
decreases at the rate
$2\gdep\sum_i\langle\nhat_i\rangle(1-\langle\nhat_i\rangle)$, so the integral
$K$ of the single-site variance measures the loss of purity, the counterpart
of $B_j$ for the state rather than for the error. Across the 122 instances it is nearly uncorrelated with $B_j$. The relative scatter of $B_j$ about
a fit in $K$ is 29.6\% against 30.2\% about its mean ($r=-0.15$). The variance
of $\alpha$ over the final distribution ($r=+0.42$, 21.2\%) and the MIS
degeneracy $g$ ($r=-0.55$, 21.6\%) are more strongly correlated with it [Fig.~\ref{fig:mechanism}], for
the reason given by Eq.~\eqref{eq:ddepdecomp}.

\begin{figure}[t]
\centering
\includegraphics[width=\columnwidth]{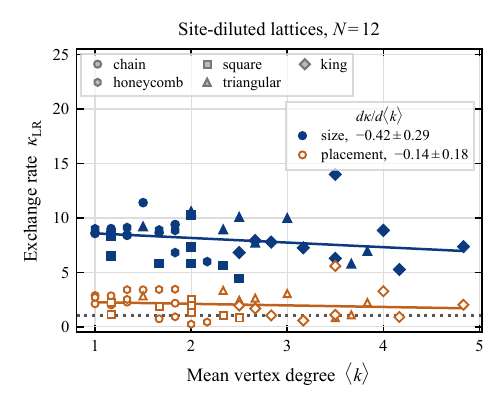}
\caption{\textbf{Degree dependence.} First-order rates at $t_f=4.56~\mu$s
against realized mean degree for the 42 site-diluted lattices at $N=12$;
marker shape, lattice family; filled, raw size error; open, raw placement
error; lines, linear fits.}
\label{fig:degree}
\end{figure}

\emph{Connectivity.} Site-diluted lattices retain each site with probability
$p$, the parent lattice being enlarged as $p$ falls so that the atom number
stays fixed; spacings are 0.85 (chain, honeycomb, square, triangular) and 0.62
(king's). A joint fit with one intercept per family,
\begin{equation}
  \log_{10}\kap_j=\mu_{\rm fam}+a\log_{10}g+b\,\langle k\rangle ,
  \label{eq:condfit}
\end{equation}
gives $a=+0.19\pm0.02$ and $b=-0.006\pm0.013$ on the 42 lattices, with
degree and degeneracy anticorrelated ($r=-0.21$), so the degree enters the rate only through the degeneracy [Fig.~\ref{fig:degree}].
Table~\ref{tab:specials} lists the engineered graphs; $K_{2,4}$ is omitted
because the schedule fails to prepare its maximum set ($\alpha_0=0.76$).

\begin{table}[t]
\caption{Engineered graphs: atom number, mean degree, MIS degeneracy and raw
size-error rate at $t_f=4.56~\mu$s. WC marks well-covered graphs, on which the
placement error vanishes.}
\label{tab:specials}
\begin{ruledtabular}
\begin{tabular}{lccccc}
family & $N$ & $\langle k\rangle$ & $g$ & $\kap_{j}$ & WC \\
\hline
cycles $C_n$, even $n$ & 4--8 & 2 & 2 & 7.5--7.9 & $C_4$ \\
cycles $C_n$, odd $n$ & 3--7 & 2 & $n$ & 14.3--15.9 & yes \\
complete $K_n$ & 4--8 & $n{-}1$ & $n$ & 17.4--26.1 & yes \\
$K_4{-}e$, $K_5{-}e$ & 4, 5 & 2.5, 3.6 & 1 & 7.1, 4.4 & no \\
stars $K_{1,n}$ & 4--6 & 1.5--1.7 & 1 & 7.2--3.1 & no \\
bipartite $K_{2,3}$ & 5 & 2.4 & 1 & 3.43 & no \\
paths $P_n$, odd $n$ & 5, 7 & 1.6, 1.7 & 1 & 7.6, 7.8 & no \\
paths $P_n$, even $n$ & 4--8 & 1.5--1.8 & 3--5 & 10.5--11.8 & $P_4$ \\
matchings $kK_2$ & 4--8 & 1 & $2^k$ & 11.9 & yes \\
two triangles $2K_3$ & 6 & 2 & 9 & 14.8 & yes \\
wheels $W_5$, $W_6$ & 6, 7 & 3.3, 3.4 & 5, 2 & 13.2, 7.2 & no \\
ladders $L_3$, $L_4$ & 6, 8 & 2.3, 2.5 & 2 & 7.0, 6.7 & no \\
tetrahedron & 4 & 3 & 4 & 17.4 & yes \\
octahedron & 6 & 4 & 3 & 8.0 & yes \\
cube $Q_3$ & 8 & 3 & 2 & 5.6 & no \\
prisms $K_2{\times}C_3$, $K_2{\times}C_5$ & 6, 10 & 3 & 6, 10 & 13.0, 11.0 & yes \\
icosahedron & 12 & 5 & 20 & 20.2 & no \\
empty $E_6$ & 6 & 0 & 1 & 8.5 & yes \\
\end{tabular}
\end{ruledtabular}
\end{table}

\section{Terminal deletion responses}
\label{app:deletion}
\emph{General form.} At $t_f$, $\hat A_f=-\sum_S\epsilon(S)\ket{S}\bra{S}$.
With $\ket{\psi}=\sum_S\psi_S\ket{S}$ and $p_0(S)=|\psi_S|^2$,
$\hat L_1^{(i)}\ket{\psi}=\sum_{S\ni i}\psi_S\ket{S\setminus i}$; since
$S\mapsto S\setminus i$ is injective on sets containing $i$, the jump term of
Eq.~\eqref{eq:damage} is $-\sum_{S\ni i}p_0(S)\epsilon(S\setminus i)$ and the
no-jump term $-\sum_{S\ni i}p_0(S)\epsilon(S)$, which gives
Eq.~\eqref{eq:delresp}. For dephasing,
$\nhat_i\ket{\psi}=\sum_{S\ni i}\psi_S\ket{S}$, the two terms coincide and
$D_{\rm dep}(t_f)=0$. The argument holds for any readout, since a readout only
changes $\epsilon(S)$.

\emph{Size error.} Each removal raises $j$ by one, so each bracket in
Eq.~\eqref{eq:delresp} is $1/|S^*|$, the inner sum is $\alpha(S)$, and the mean
is $\alpha_0$.

\emph{Placement error.} All maximum sets have the same size, so
the distance $|S\setminus S^*|+|S^*\setminus S|=|S|+|S^*|-2|S\cap S^*|$ has the same parity for every
$S^*\in\mathcal{M}$. Removing $i$ changes this distance by $+1$ if
$i\in S^*$ and by $-1$ otherwise. If some nearest $S^*$ omits $i$, then
$d(S\setminus i)=d(S)-1$, since no removal lowers a distance by more than one.
If every nearest $S^*$ contains $i$, these give $d(S)+1$, and every other
$S^*$, at distance at least $d(S)+2$ by parity, gives at least $d(S)+1$; so
$d(S\setminus i)=d(S)+1$. As $j$ rises by one at every removal and $d=j+2r$,
the first case lowers $r$ by one and the second leaves it unchanged. Summing,
$\sum_{i\in S}[r(S\setminus i)-r(S)]=-w(S)$, which gives
Eq.~\eqref{eq:termjr}. A deletion never increases the placement error.

\emph{Post-processed size error.} Let $S$ be maximal and $i\in S$. A vertex
$v\notin N[i]$ outside $S\setminus i$ is outside $S$, so it has a neighbor
$u\in S$ with $u\neq i$ and is blocked in $S\setminus i$; the free vertices of
$S\setminus i$ therefore lie in the closed neighborhood $N[i]$, and $i$ is one
of them. Greedy completion returns $(S\setminus i)\cup T$ with
$T\subseteq N[i]$ nonempty, so $|\Pg(S\setminus i)|\ge|S|$. A swap never
lowers the size, so $|\Ps(S\setminus i)|\ge|\Pg(S\setminus i)|\ge|S|$. If $S$
is a fixed point of the readout, $\PP(S)=S$ and each bracket for the
post-processed size error is zero or negative. For outputs that the readout
changes, greedy completion is not monotone and the bracket has no definite
sign, so only such outputs can give a positive terminal deletion response.

\section{Quasistatic detuning noise}
\label{app:classical}
A quasistatic detuning offset $\epsilon_i\sim\mathcal{N}(0,\sigma^2)$, drawn once per
shot and constant during it, enters as an extra phase and costs one noiseless propagation
per realization. Its first-order effect averages to zero, leaving
$\bar{\alpha}=\alpha_0-C\sigma^2$ with $C=0.0011\pm0.0003~\mu$s$^2$ (antithetic
pairs, 128 per instance; valid to 30\% for $\sigma=1$--$5~\mu$s$^{-1}$). At the
calibrated Doppler width $\sigma/2\pi=50$~kHz~\cite{Dalyac2026} the loss is
$1.1\times10^{-4}$, or 0.4\% of the Markovian size error added at FC1.
Recasting the spread as a Lindblad rate $\sigma^2t_f\approx0.4~\mu$s$^{-1}$
would instead make it comparable to the edge of our grid. The recast fails for the
reason known from filter-function
analyses~\cite{Cywinski2008,Bylander2011,Green2013}: a static offset is a
rigid shift of the sweep, which the adiabatic anneal follows.

\end{document}